\documentclass[12pt]{article}

\usepackage[a4paper, margin=1in]{geometry}
\usepackage{amsmath,bm}
\usepackage{amssymb}
\usepackage{amsfonts}
\usepackage{commath}
\usepackage{graphicx}
\usepackage{subcaption}
\usepackage{siunitx}
\usepackage{enumerate}
\usepackage{fancyhdr}
\usepackage{fancyvrb}
\usepackage{relsize}
\usepackage{lastpage}
\usepackage{pdfpages}
\usepackage{multirow}
\usepackage{svg}
\usepackage{float}
\usepackage{array}
\usepackage{here}
\usepackage{lineno}
\usepackage{sectsty}
\usepackage{gnuplottex}
\usepackage{url}
\usepackage{hyperref}
\usepackage{authblk}
\usepackage{caption}
\usepackage{fancyvrb}
\usepackage{relsize}
\usepackage{wrapfig}
\usepackage{setspace}
\usepackage[numbers]{natbib}
\usepackage{lineno}
\usepackage{authblk}

\usepackage[font=scriptsize]{caption}

\title{The Fluid Mechanics of Truncus Arteriosus}
\date{}

\author[1,2]{Karoline-Marie Bornemann, PhD}
\author[2,3,4]{Perry S. Choi, MD}
\author[3]{Jay A. Huber, MAB}
\author[3,4]{Hannah L. McMullen, MD}
\author[3,4]{Alexander K. Reed, MD}
\author[3,4]{Amit Sharir, BS}
\author[1,2,4]{Shiraz A. Maskatia, MD}
\author[1,2,4,5,6]{Alison L. Marsden, PhD}
\author[2,3,4]{Michael R. Ma, MD}
\author[2,3,4,$\ast$]{Alexander D. Kaiser, PhD}

\affil[1]{Department of Pediatrics (Cardiology), Stanford University, Stanford, CA 94305, USA}
\affil[2]{Maternal \& Child Health Research Institute, Stanford University, Stanford, CA 94305, USA}
\affil[3]{Department of Cardiothoracic Surgery, Stanford University, Stanford, CA 94305, USA}
\affil[4]{Cardiovascular Institute, Stanford University, Stanford, CA 94305, USA}
\affil[5]{Department of Bioengineering, Stanford University, Stanford, CA 94305, USA}
\affil[6]{Institute for Computational \& Mathematical Engineering, Stanford University, Stanford, CA 94305, USA}

\begin{document}

\maketitle
\thispagestyle{empty}

\subsubsection*{Corresponding Author Address}
Alexander D. Kaiser, PhD \\
Department of Cardiothoracic Surgery, Stanford University \\
Clark Center, E100, 318 Campus Dr, Stanford, CA 94305, USA \\
E-mail: alexdkaiser@stanford.edu \\

\subsubsection*{Key terms} 
Fluid-structure interactions; Congenital heart disease; Neonatal cardiac surgery; Vortex dynamics; Lagrangian Coherent Structures; Total mixing lesion; Truncus arteriosus

\newpage
\setcounter{page}{1}

\subsubsection*{Abstract} 
\textbf{Background:} Truncus arteriosus (TA) is a rare and severe congenital heart disease in which the two main arteries exiting the heart fail to separate in utero resulting in one truncus and truncal valve, that carries mixed oxygenated and deoxygenated blood. Approximately 25\% of patients present with a quadricuspid valve, which is prone to regurgitation and re-intervention. Despite the relevance of the interaction between the truncal valve and blood flow for valve performance and mixing, fluid mechanics in TA are poorly understood and understudied. \\
\textbf{Methods:} Patient-specific fluid-structure interaction simulations were performed based on computed tomography imaging segmentation before and after TA repair. The quadricuspid valve was constructed based on elasticity-based design with the patient’s free edge length and geometric height extracted from echocardiography. The interaction between blood and valve was simulated with the Immersed Boundary Method. Boundary conditions were tuned based on the patient’s data. Lagrangian Coherent Structures (LCS) and Lagrangian Particle Tracing (LPT) were used to investigate mixing of oxygenated and deoxygenated blood and streaming. \\
\textbf{Results:} Low pressures, forward flow, and streamwise vortices in the one-sided PAs during the entire cardiac cycle affected the leaflet motion, leading to asymmetric closure and regurgitation. Holodiastolic aortic flow reversal supplied PA flow and the regurgitant jet. LCS and LPT indicated favorable streaming of oxygenated blood from the LV into the aorta and deoxygenated blood from the RV into the PAs. After truncal surgery, normal hemodynamics were restored. \\
\textbf{Conclusion:} This is the first study of fluid mechanics of TA. Using qualitative and quantitative methods of flow analysis in patient-specific models, we identified disrupted preoperative hemodynamics caused by one-sided PA origin and showed how normal hemodynamics were re-established after repair. Favorable streaming was demonstrated aligned with patient’s reports. Thus, favorable streaming is a plausible mechanism in total mixing lesions and patient-specific modeling may aid in its detection.

\newpage

\section{Introduction} 
\label{sec:introduction}

Truncus arteriosus (TA) is a rare and severe congenital heart defect that occurs when the two main arteries exiting the heart fail to separate in utero. TA requires surgical repair within the first few weeks after birth. In TA, the heart has a single truncus with one outlet valve, the truncal valve, that carries mixed deoxygenated and oxygenated blood. A ventricular septal defect (VSD) allows blood flow to enter from both left ventricle (LV) and right ventricle (RV) upstream of the truncal valve. Downstream of the valve, the pulmonary arteries (PAs) branch from the aorta. The flow entering the PAs is often more than three times as high as the flow entering the aorta, resulting in pulmonary over-circulation \cite{Naimo2021, Martinez2019}. 

Truncal repair includes the closure of the VSD, detachment of the PAs from the aorta, and reconnection of the right ventricle to the PAs via a valved RV-PA conduit. The most relevant determinant of success in TA repair is the postoperative competency of the truncal valve, with proper valve opening during systole allowing unobstructed outflow and tight closure during diastole preventing backflow into the ventricle \cite{Naimo2021}. Commonly, the truncal valve leaflets are thickened and dysplastic, showing limited mobility. While the majority of patients with TA presents with a three-leaflet valve, quadricuspid valve morphology occurs in around 25\% of all TA patients \cite{Guariento2022}, which is categorized into 7 phenotypes \cite{Hurwitz1973}. Over one third of truncal patients with quadricuspid valve morphology require re-intervention on the valve itself with tricuspidization showing the longest durability among surgical strategies \cite{Naimo2021b}. The most prominent failure mode of quadricuspid valves is regurgitation, which can be further aggravated by progressive annular dilation, necessitating re-intervention during early childhood \cite{Konstantinov2023}. 

Although TA has been evaluated by echocardiography imaging for Doppler-based flow quantification (peak velocities, pressure gradients, regurgitation) \cite{Martinez2019, Naimo2021, Chen2016, Naimo2018}, TA remains highly understudied from an engineering perspective, particularly combined with concomitant quadricuspid valve morphology. Williams and colleagues \cite{Williams2022} used image-derived models to inform truncal valve repair in an adolescent patient with previous TA repair and diagnosed with progressive valve insufficiency and quadricuspid valve detected during postoperative serial imaging. Using the SlicerHeart extension of 3DSlicer \cite{Lasso2022}, the authors performed volume rendering and segmentation of the quadricuspid valve to identify bicuspidization as the optimal repair. However, no fluid or structural mechanics were simulated. Vismara and colleagues \cite{Vismara2014} performed an in vitro study of porcine quadricuspid valves in a pulsatile mock circulatory flow loop. Using high-speed camera measurements, the authors identified longer opening and closing times and an asynchronous closing compared to the three-leaflet control valve while hemodynamics remained similar. Choi and colleagues \cite{Choi2025} created a diseased, porcine model of a quadricuspid valve and evaluated the impact of no-cut tricuspidization and bicuspidization repairs. While both repair techniques showed improvement in regurgitation and stenosis, tricuspidization performed better than bicuspidization. 

In contrast to quadricuspid valve morphology, hemodynamics and leaflet kinematics have been extensively studied in three-leaflet native aortic valve morphology \cite{Kaiser2021, Fringand2024, Gilmanov2018, Chen2020} and their prosthetic replacements \citep{Bornemann2026_JFM, Ferrari2024, Lee2020, Oechtering2019, Johnson2022, Griffith2012} as well as congenital bicuspid valves \cite{Kaiser2022, Saikrishnan2012, Marom2013b, Lavon2018} and their repair \cite{Kaiser2024, Kaiser2025, Choi2024}. When the ventricular pressure exceeds the aortic pressure, the valve opens and a starting vortex is formed. The shape and coherence of this vortex are highly dependent on the initial valve orifice and hemodynamic conditions \cite{Bornemann2026_JFM}. During systolic acceleration, the aortic jet enters the ascending aorta and impinges at the outer aortic curvature, leading to retrograde flow along the aortic wall toward the sinuses of Valsalva. Instabilities in the shear layer lead to the transition from laminar to turbulent flow with their onset dependent on valve geometry, valve kinematics, and surrounding patient anatomy \cite{Bornemann2024_JFM, Bornemann2025_POF}. During systolic deceleration, the turbulent aortic jet decelerates, turbulent flow dissipates and the valve closes when the aortic pressure exceeds the ventricular pressure. Ideally during diastole, the flow motion in the ascending aorta is negligible and valve regurgitation is absent. 

Besides the understudied fluid-structure interaction in quadricuspid valve morphology, the complex anatomy of TA with two ventricular inlets and three downstream outlets further complicates the blood flow motion. To this end, no study on the fluid mechanics in TA exists in literature, although a comprehensive understanding would assist in predicting valve performance and mixing patterns. TA is classified as total mixing or admixture lesion, as complete mixing of oxygenated and deoxygenated blood is possible at the level of VSD, truncal valve and in the truncus. Against the general assumption that total mixing occurs in TA, some existing literature points to the possibility of favorable or preferential streaming in admixture lesions \cite{Kulik2017} comparable to the well-established concept of streaming in the fetal circulation \cite{Schrauben2019}. Among complex CHD, qualitative assessment of streaming was only evaluated for transposition of the great arteries \cite{Wong2014} to the authors' knowledge. Wong and colleagues \cite{Wong2014} identified very little mixing between the two ventricles based on particle tracing in 4D Cardiac Magnetic Resonance Imaging leading to unfavorable streaming conditions and low oxygen saturation in the investigated patient. While this shows how the analysis of streaming patterns in children with congenital heart disease would be highly informative to assess mixing of oxygenated and deoxygenated blood, no studies analyzed the fluid mechanics in TA to quantify streaming patterns in postnatal, diseased anatomy. 

\begin{figure}[hp!]
    \centering
    \includegraphics[width=\textwidth]{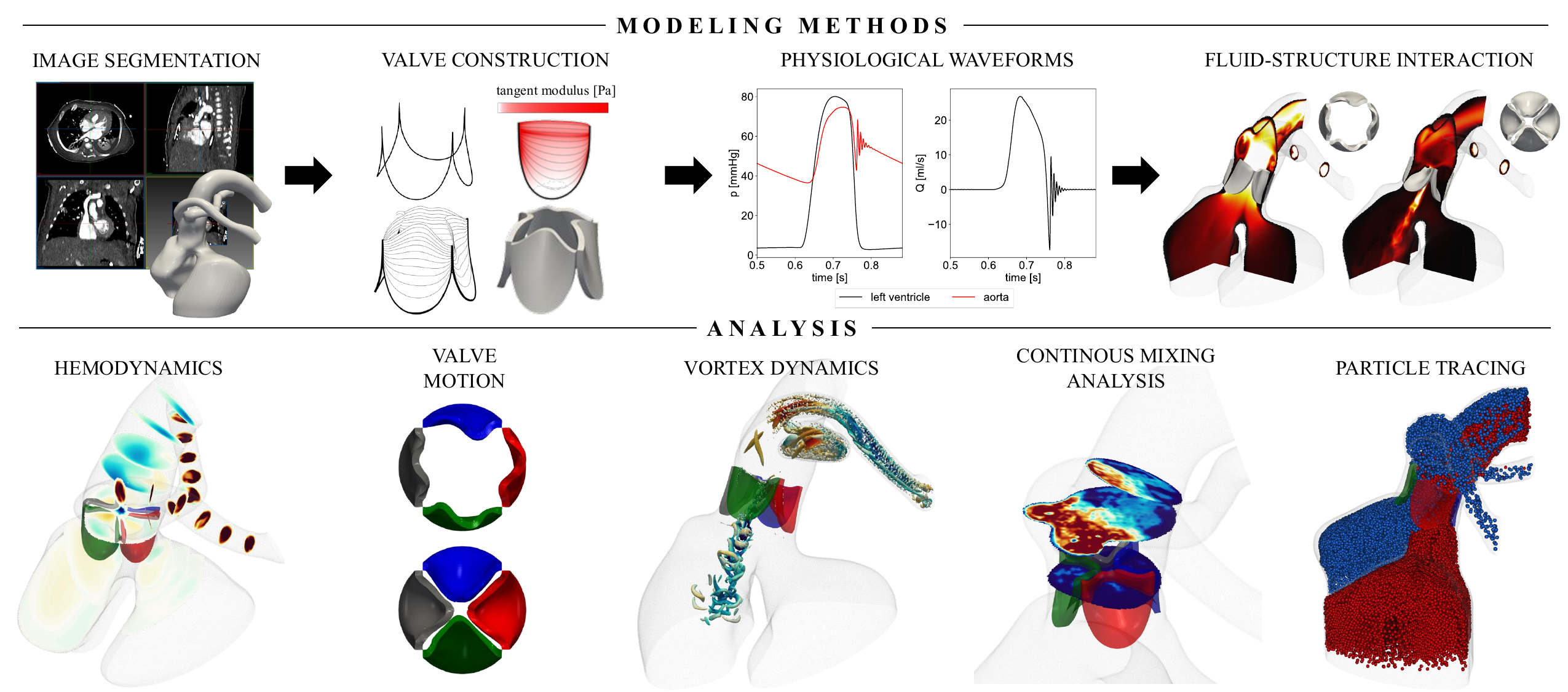}
    \caption{\textbf{Overview of modeling methods and analysis performed in this study.} Patient-specific modeling methods: (1) Image segmentation, (2) Valve construction, (3) Physiological waveforms and (4) Fluid-structure interaction. Analysis performed in this study to assess fluid mechanics of truncus arteriosus: (1) Hemodynamics, (2) Valve motion, (3) Vortex dynamics, (4) Continuous mixing analysis (Lagrangian Coherent Structures) and (5) Lagrangian Particle Tracing.}
    \label{fig:workflow}
\end{figure}

This study is the first investigation of the fluid mechanics of TA. Using qualitative and quantitative methods of flow analysis in patient-specific models, we identified disrupted hemodynamics and favorable mixing patterns in the diseased, preoperative state of TA and showed how normal hemodynamics are re-established through its surgical repair.
Previously, we evaluated the clinical implications of hemodynamics on valve performance in a patient before and after truncal repair without direct leaflet intervention \cite{Bornemann2026_truncal}. We showed that our computational simulations can reproduce phenotypes that change with repair and identify mechanisms improving postoperative valve performance, demonstrating the viability and clinical relevance of our approach. In the present study, we utilized tools of classical flow analysis to understand the fluid mechanics in TA and relate them to mixing patterns of oxygenated and deoxygenated blood (Figure \ref{fig:workflow}). Modeling the patient-specific fluid-structure interaction between blood flow, truncal valve and vessel anatomy, we showed abnormal fluid mechanics in the preoperative state and how surgical repair re-establishes fluid mechanics comparable to that of a healthy heart.

\section{Methods}
\label{sec:methods}

The interaction of valve leaflets and vessel wall with the surrounding blood flow was modeled via the Immersed Boundary Method implemented in the software library IBAMR \cite{Griffith2007}. The fluid was modeled as incompressible with a density of $\rho = 1.0 \, g/cm^3$ and a dynamic viscosity of $\mu = 0.04\, Poise$ based on the Navier-Stokes equations including a structure force term acting on the fluid. Three cycles are simulated of which the second cycle is used for analysis. 

The neonatal patient was selected based on the availability of pre- and postoperative imaging and the absence of any surgical modification of the truncal valve. The patient was diagnosed with quadricuspid valve morphology type A \cite{Hurwitz1973} and truncus arteriosus type II \cite{Collett1949}. Segmentation of CT imaging scans was performed in the open-source software SimVascular \cite{Simvascular2016} to create pre- and postoperative vessel geometries (Figure \ref{fig:methods}A). The quadricuspid valve was constructed based on an elasticity-based design approach introduced by Kaiser and colleagues \cite{Kaiser2019, Kaiser2021}. Using this methodology, the heterogeneous leaflet fiber structure and material stiffnesses were derived via tuning parameters in the leaflet equilibrium equations under the condition that the valve supports a given pressure (Figure \ref{fig:methods}B). Patient-specific free edge length (FEL) and geometric height (GH) were measured from echocardiography imaging before and after TA repair (Figure \ref{fig:methods}C). For the preoperative configuration, we matched the gross valve morphology to a FEL of $7.56\,mm$ and GH of $6.5\,mm$ as well as a FEL of $8.95\,mm$ and GH of $6.5\,mm$ for the postoperative configuration via elasticity-based design \cite{Kaiser2021}. Patient-specific boundary conditions were derived from clinically measured patient vitals, CT and echocardiography imaging and tuned based on multiple iterations between high-fidelity, three-dimensional FSI simulations and the equivalent zero-dimensional configuration using svZeroDSolver implemented in SimVascular \cite{Pfaller2021} (Figure \ref{fig:methods}C).

To assess the flow field in pre- and postoperative configuration, vessel centerlines were extracted and the streamwise velocity component $\mathbf{u}_s$ was computed. Flow and pressure waveforms were extracted at the inlets (LV, RV) and outlets (aorta, LPA, RPA) as well as at a slice located $0.12$ cm above the valve commissure level which corresponds to the region upstream of the PAs in the preoperative configuration. To evaluate leaflet kinematics, the location of the central leaflet tip was tracked over time and its Euclidean distance to the central leaflet coaptation point during diastole was calculated. Vortex development and breakdown were assessed by iso-surfaces of the Q-criterion. Lagrangian Coherent Structures were visualized via ridges of the Finite Time Lyapunov Exponent (FTLE) field, identifying transport barriers in the flow field \cite{Shadden2005}. Mixing of oxygenated and deoxygenated blood was qualitatively assessed via massless Lagrangian particles, which were seeded within the ventricles. Particles within the LV were assumed as `oxygenated' (colored red) and particles within the RV were assumed as `deoxygenated' (colored blue). Particles were released just prior to the valve opening and traced over the second cardiac cycle.

The Institutional Review Board of the Stanford University approved the study protocol and publication of data. Patient written consent for the publication of the study data was waived by the Institutional Review Board for use of anonymized, retrospectively acquired data (\#39377, June 17, 2025).

More details of the methods were described in the Supplementary Material.

\section{Results}
\label{sec:results}

\subsection{Hemodynamics, waveforms and leaflet kinematics}
\label{subsec:hemodynamics}

Key elements of the preoperative patient-specific anatomy included left-sided anterior closely-spaced origins of both branch PAs from the TA. The leaflets aligned beneath (upstream of/proximal to) the PA origins were the left and anterior leaflets while the right and posterior leaflets were opposite.

The one-sided location of the PAs and their associated lower pressure levels of the pulmonary circulation led to abnormal hemodynamics in the preoperative state, while expected hemodynamics were re-established through surgical repair of TA. As the PAs acted as a pressure sink, the connection of the systemic and pulmonary circulation led to abnormally low pressure levels immediately downstream of the valve throughout the cardiac cycle. Valve regurgitation and forward flow in the PAs were both supplied by holodiastolic flow from the aorta. Truncal surgery restored expected hemodynamics such as a forward aortic jet, normal pressures downstream of the valve and substantially reduced valve regurgitation.

\begin{figure}
    \centering
    \includegraphics[width=\textwidth]{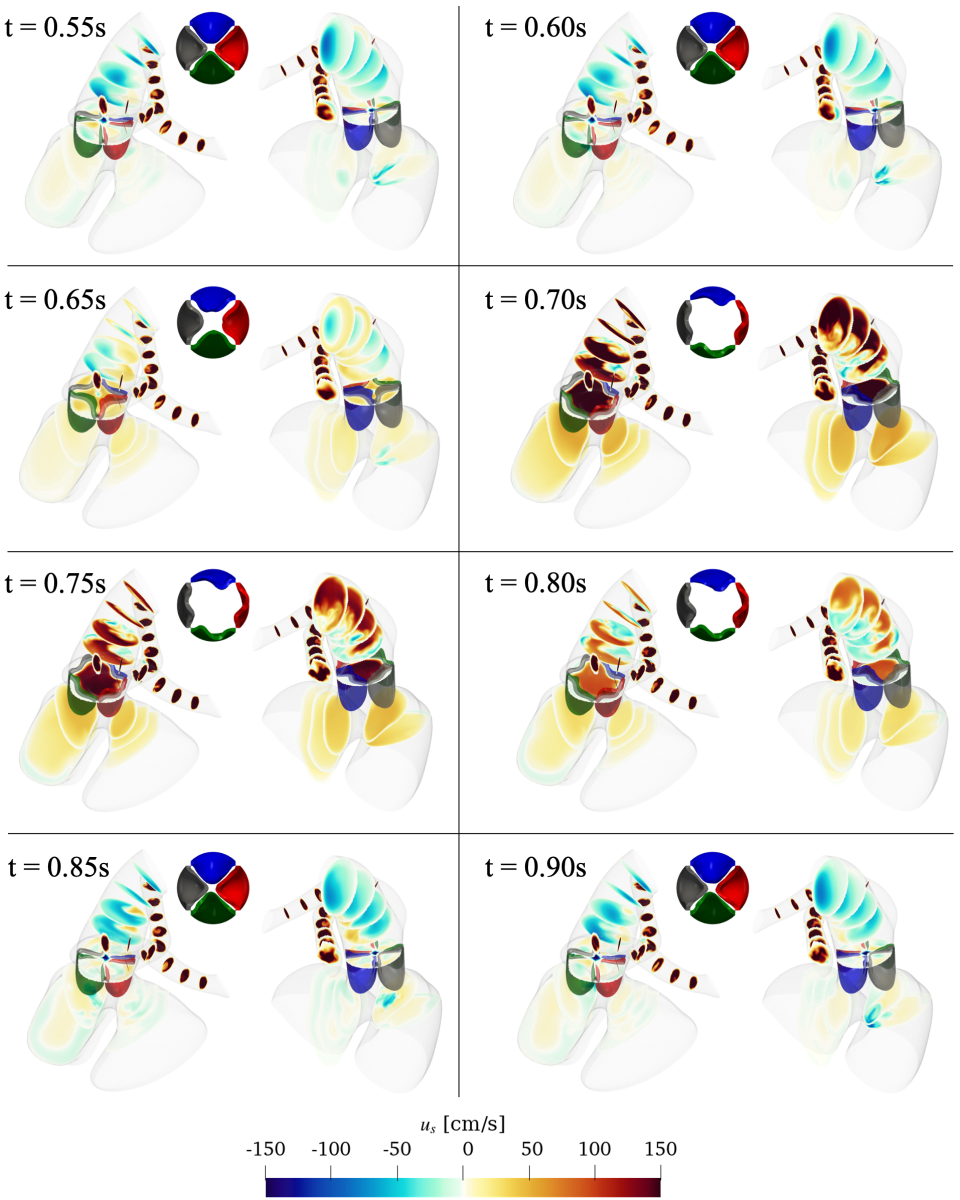}
    \caption{Preoperative configuration: Streamwise velocity component $u_s$ tangent to the respective vessel centerlines and top view of valve across one cardiac cycle ($t = [0.55,0.90]s$). Both forward flow in the pulmonary arteries during the entire cardiac cycle and valve regurgitation were supplied by holodiastolic flow reversal in the aorta. The regurgitant jet through the truncal valve entered the right ventricle.}
    \label{fig:hemodynamics_preop}
\end{figure}

Evaluating the flow along the vessel centerlines, an asymmetric truncal jet inclined towards the outer curvature of the truncus originating from the asymmetric valve orifice, combined with retrograde flow along the inner truncus curvature (Figure \ref{fig:hemodynamics_preop}) during systole. Compared to the aorta, streamwise velocities were substantially higher in the PAs during systole. During asymmetric valve closure, the truncal jet slowed down and flow direction in the truncus reversed, while forward flow in the PAs remained throughout diastole. During the entire diastolic phase, we observed retrograde flow from the aorta feeding the narrow regurgitant jet through the valve and into the RV. 
 
\begin{figure}
    \centering
    \includegraphics[width=\textwidth]{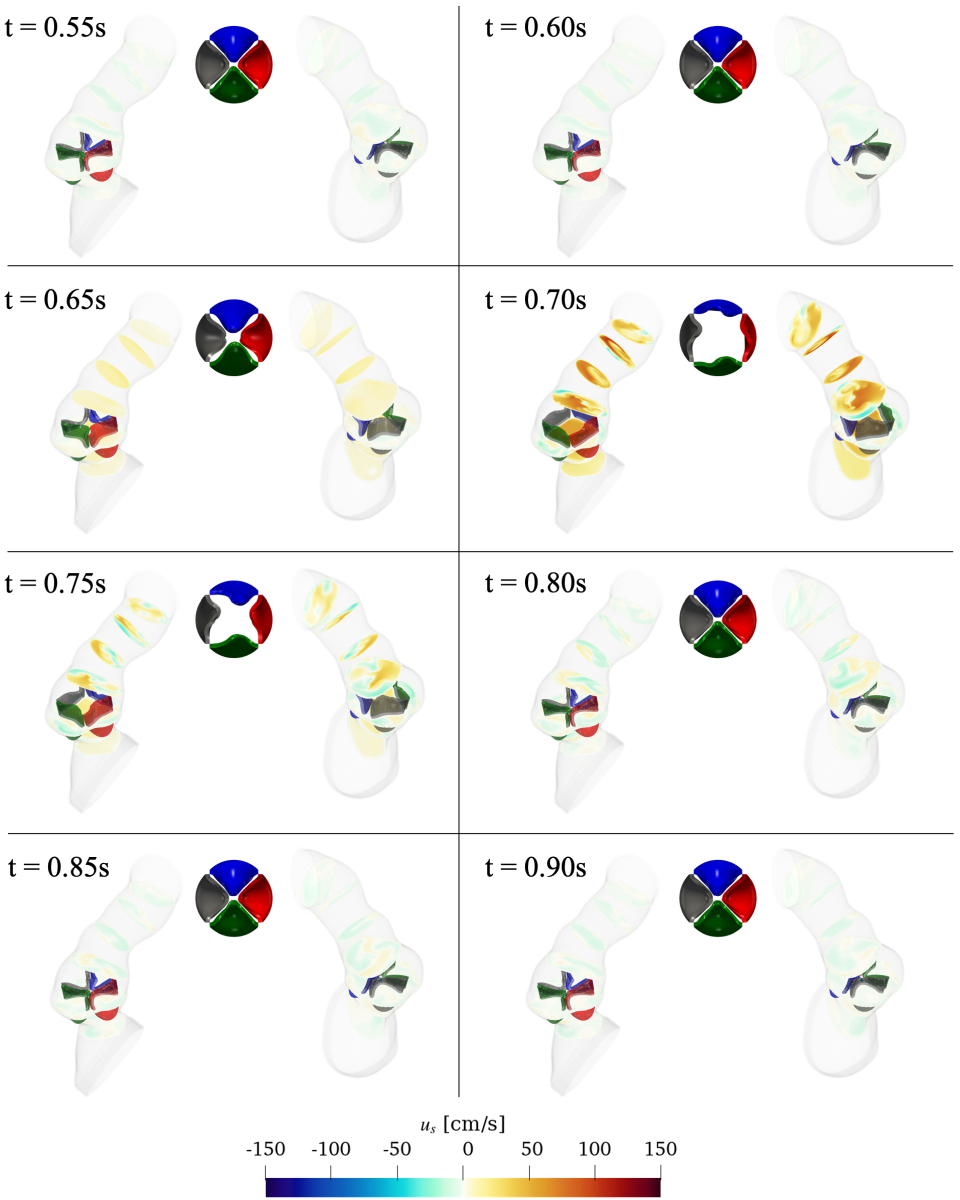}
    \caption{Postoperative configuration: Streamwise velocity component $u_s$ tangent to the respective vessel centerlines and top view of valve across one cardiac cycle ($t = [0.55,0.90]s$). Truncal surgery restored expected hemodynamics such as a forward aortic jet, normal pressures downstream of the valve and substantially reduced valve regurgitation.}
    \label{fig:hemodynamics_postop}
\end{figure}

Compared to the preoperative configuration, we observed significantly less flow motion during diastole in the postoperative configuration (Figure \ref{fig:hemodynamics_postop}). During valve opening, skewed LV inflow resulted in an inclined aortic jet towards the outer vessel curvature. During systole, peak streamwise velocities were lower compared to the preoperative configuration. During valve closure, the aortic jet dissipated towards almost zero flow motion during diastole. Holodiastolic flow reversal was eliminated after truncal repair.

\begin{figure}
    \centering
    \includegraphics[width=\textwidth]{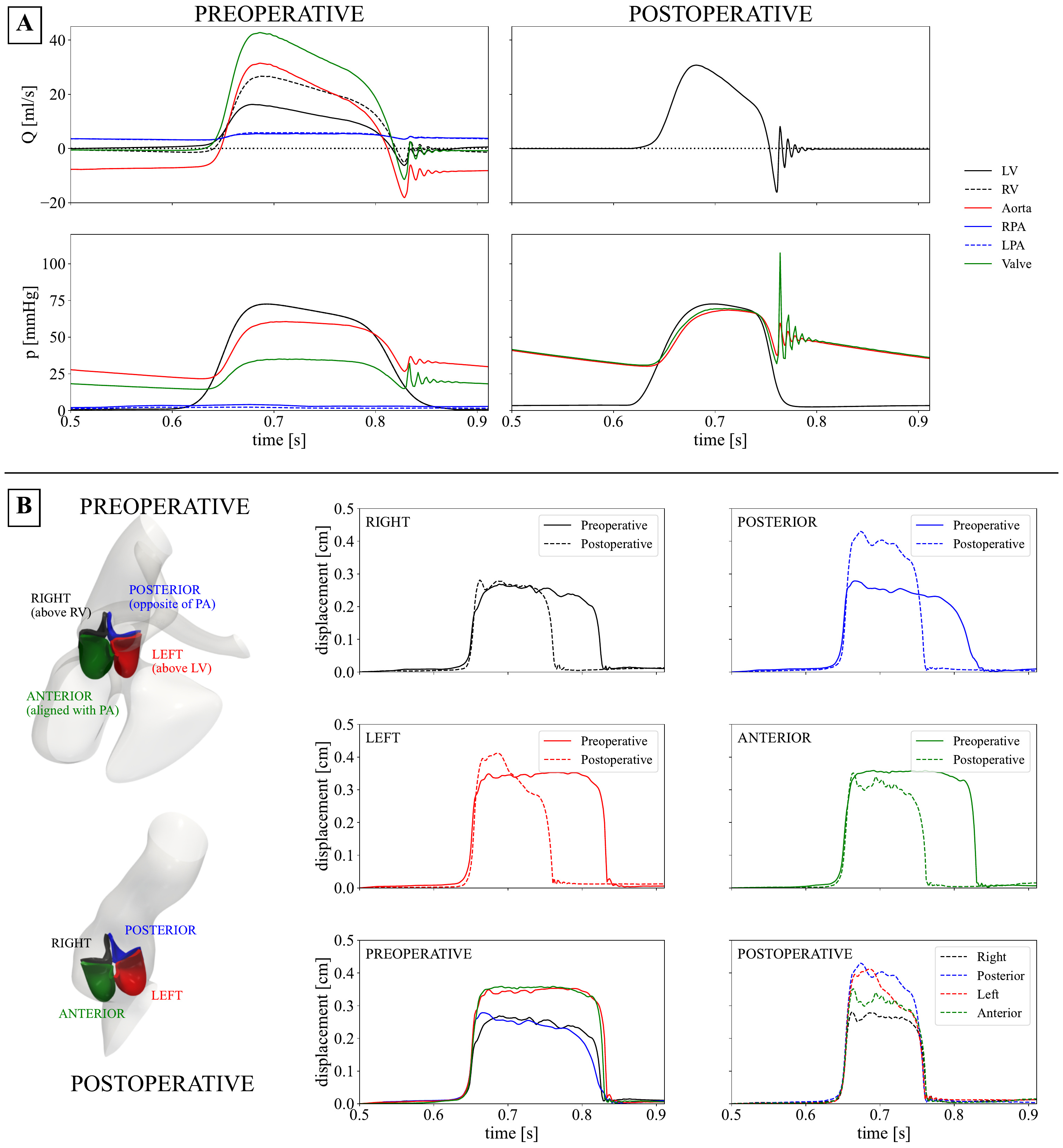}
    \caption{(\textbf{A}) Pre- and postoperative pressure and flow waveforms spatially averaged over a slice at the inlets (left ventricle, right ventricle) and the outlets (aorta, left and right pulmonary artery). The green waveform (`Valve') describes the spatially averaged flow and pressure immediately distal to the valve, which is computed at a slice located $0.12$ cm above the valve commissure level (below the pulmonary artery origin for the preoperative configuration). As the pulmonary arteries acted as a pressure sink, pressure levels immediately downstream of the valve are lower in preoperative compared to postoperative configuration. (\textbf{B}) Leaflet kinematics over one cardiac cycle for pre- and postoperative configurations. Interaction between the heterogeneous downstream flow field created by the one-sided location of the PAs and the truncal valve led to asymmetric valve closure. Leaflets opened and closed in pairs with the leaflets upstream of the PAs opening wider and closing later. Restoring healthy hemodynamics by truncal repair resulted in more symmetric valve closure.}
    \label{fig:waveforms_kinematics}
\end{figure}

Pre- and postoperative pressure and flow waveforms at the inlets (LV, RV) and outlets (aorta, LPA, RPA) are shown in Figure \ref{fig:waveforms_kinematics}A. In the preoperative configuration, we observed a $25\%$ higher stroke volume contribution from the RV compared to the LV and higher peak flow rates (LV: $16.25\, ml/s$, RV: $26.67 \, ml/s$) during systole. The combined flow from the ventricles entered the valve, leading to a total stroke volume minus the regurgitant volume of $4.87\,ml$. The regurgitant flow through the truncal valve mostly entered into the RV during diastole. The peak flow rate downstream of the valve was $42.76 \, ml/s$ during systole. Flow waveforms showed relatively constant forward flow in the PAs during the entire cardiac cycle. The regurgitant jet through the valve during diastole was fed entirely by holodiastolic retrograde flow in the aorta, with no retrograde diastolic flow contribution from the branch PAs. The peak local preoperative Reynolds number at the valve annular plane based on the annular diameter and the spatially averaged velocity magnitude was $1721$. In the postoperative configuration, the peak flow rate during systole was $31.14\,ml/s$ and substantially reduced regurgitation was observed. The peak local Reynolds number at the valve inflow was $618$.

The significant difference between aortic (systemic circulation) and PA (pulmonary circulation) pressures contributed to elevated flow towards the pulmonary circulation with a flow ratio of $3.2$ to $1$. While postoperative pressures just downstream of the aortic valve were ranging between ventricular and aortic pressures due to viscous losses, pressure levels downstream of the truncal valve were unexpectedly low in the preoperative configuration. Instead of slightly higher pressures compared to the aorta, the truncal pressure was substantially lower and ranged between the aortic and pulmonary pressure during the entire cardiac cycle. These abnormally low pressure levels were caused by the connection of the higher-pressure systemic circulation and lower-pressure pulmonary circulation.  

The interaction between the heterogeneous downstream flow field created by the one-sided location of the PAs and the truncal valve consequently led to asymmetric valve closure in the preoperative configuration (Figure \ref{fig:waveforms_kinematics}B). Preoperative leaflets opened and closed in pairs, with the left and anterior leaflets distal to the left-sided anterior PA orifices opening wider and closing later. The left anterior leaflets distal to the left-sided anterior PAs showed higher maximum displacement during systole (left: $3.6\,mm$, anterior: $3.7\,mm$) and later closing. Posterior and right leaflets, located opposite the PA origins, showed lower maximum displacements during systole of (posterior: $3.0\,mm$, right: $2.7\,mm$). While opening occurred approximately simultaneously, the leaflets did not close at the same time. The posterior leaflet showed the most gradual closing slope starting at the earliest time, while the right, anterior and then left leaflet showed similar closing slopes. During diastole, the coaptation height near the valve commissures was $20.0$\% of the leaflet geometric height. Restoring healthy hemodynamics by truncal repair resulted in more symmetric valve closure (right: $3.3\,mm$, posterior: $4.3\,mm$, left: $4.3\,mm$, anterior: $3.6\,mm$). All leaflets closed approximately at the same time and show a higher coaptation height of $30.8$\% of the leaflet geometric height. 

The truncus in the preoperative configuration was of tubular shape without pronounced sinus bulges or sinotubular junction. As a result of the limited outward space, preoperative leaflets immediately settled into their stable open position during systole. The postoperative configuration showed more pronounced sinuses, and thus the leaflets opened to their maximum displacement initially and then settled in a stable open position during systole which was reflected in an initial bump of leaflet displacement in Figure \ref{fig:waveforms_kinematics}B. 

\subsection{Vortex development and breakdown}
\label{subsec:vortex}

During systolic acceleration, the valve opened and a starting vortex was formed. The starting vortex shape highly depends on the leaflet shape, the resulting orifice shape and the hemodynamic conditions \cite{Bornemann2024_JFM, Bornemann2025_POF, Bornemann2026_JFM}. Here, quadricuspid valve morphology led to an approximately four-lobed starting vortex. However, surrounding vessel geometry highly influenced its development and deformation. In the preoperative configuration, coherent streamwise vortices in the PAs persisted throughout the entire cardiac cycle and highly influenced the flow field and vortex dynamics in the preoperative configuration (Figure \ref{fig:qcrit_FTLE_particle}A). During the initial valve opening, these vortices merged with the starting vortex. Generally, the starting vortex was less coherent initially and broke up quicker than in the postoperative configuration. During systole, small-scale vortices along the inner truncal curvature were observed accompanied by streamwise vortices in the PAs. During diastole, a transitional regurgitant jet was directed toward the RV while the streamwise vortices in the PAs persisted. Postoperatively, after removal of the PAs, three sides of the vortex remained roughly parallel to the valve orifice, while the fourth side deflected along the inner aortic curvature resulting in an elongated hairpin vortex due to immediate onset of the aortic curvature downstream of the valve. We also observed periodic vortex shedding from the inner curvature. In diastole, no coherent vortical structures were detected in the aorta, indicating a complete dissipation of the systolic turbulent flow.

\subsection{Lagrangian Coherent Structures}
\label{subsec:lcs}

Lagrangian Coherent Structures in the preoperative state revealed a distinct material surface along the VSD and an upstream effect of the PAs at valve level during systole. A material surface is defined by local maxima of the FTLE field and can be interpreted as transport barriers, across which negligible flux and therefore mixing occurs. During diastole, flow from the outer aortic curvature entered into the PAs. After truncal surgery, systolic hemodynamics were restored with a square-shaped aortic jet due to the quadricuspid valve and negligible diastolic flow motion during diastole (Figure \ref{fig:qcrit_FTLE_particle}B).

In the preoperative configuration, a material surface defined by local maxima of the FTLE field was located across the VSD and throughout the truncal valve during systole. Visualizing the FTLE field across slices normal to the truncal centerline revealed that this material surface separates the truncal jet from the branching flow into the PAs. The effect of the PAs extended upstream into the valve where we observed two looped material surfaces upstream of the PA origins. At the PA orifices, chaotic PA inflow was present next to a material surface enclosing the truncal jet. The flow through the valve appeared to be divided into a contribution originating from the RV directed towards the PAs and a truncal jet towards the outer vessel curvature. During diastole, FTLE fields demonstrated the regurgitant flow from the aorta along its outer curvature into the PAs. In the postoperative configuration, we noticed substantial differences in the absence of the PAs. FTLE fields showed the deflection of the initially square-shaped aortic jet as well as local recirculation zones forming in the irregular outer curvature created by truncal repair and two streamwise vortices along the inner curvature. Negligible flow motion existed during diastole.

\subsection{Streaming patterns of oxygenated and deoxygenated blood}
\label{subsec:streaming}

Lagrangian Particle Tracing of oxygenated particles from the LV and deoxygenated particles from the RV showed that particles from both ventricles were advected towards the valve in a helical, twisted motion (Figure \ref{fig:qcrit_FTLE_particle}C). A surface seemed to separate the distinct oxygenated and deoxygenated particle streams, respectively, until reaching the aorta. This surface aligned well with the previously determined local maxima of the FTLE field, further supporting the hypothesis of negligible mixing of oxygenated and deoxygenated blood at the VSD level during systole. As the particle stream from the RV twisted towards the PA origin, mostly deoxygenated particles entered the PAs during most of systole. In contrast, the particle stream from the LV mostly entered into the aorta. During systolic deceleration, the backflow from the aorta caused particle mixing at PA level and therefore entering of oxygenated particles into the PAs. During diastole, FTLE fields showed no distinct ridge across the VSD which agreed well with the mixing of both particle groups.

Both Lagrangian Coherent Structures and Lagrangian Particle Tracing revealed favorable strea\-ming patterns of oxygenated and deoxygenated flow with streaming from the LV into the aorta and streaming from the RV into the PAs. Preoperative clinical records of our model patient indeed noted `exceptionally high systemic oxygen saturations for a truncus arteriosus patient', with pre-operative resting oxygen saturations in the low 90s on room air. While oxygen saturations of 75\% to 85\% are often considered satisfactory goals in pre-repair total mixing lesions such as TA, this patient’s pre-operative saturation was 93\% at 21\% FiO2. This supported that we have shown, for the first time, that favorable mixing is a plausible mechanism in a total mixing lesions such as truncus arteriosus.    

\section{Discussion}
\label{sec:discussion}

This study represents the first investigation of fluid mechanics of truncus arteriosus. We showed that abnormal fluid mechanics are highly influenced by the location of the PAs, which acted as a pressure sink and therefore created an asymmetric interaction of truncal flow and the truncal valve. Disconnection of the PAs during truncal surgery re-established normal fluid mechanics and substantially reduced valve regurgitation. Furthermore, indicators for favorable flow were identified in line with the patient's clinical reports for the first time in truncus arteriosus. 

The one-sided location of the PAs and the resulting downstream hemodynamic heterogeneity interacting with the valve leaflets were the main cause of flow asymmetry. The significantly lower pressure in the pulmonary compared to the systemic circulation as well as the lower pulmonary vascular compared to systemic vascular resistance contributed to the high pulmonary to systemic flow ratio when both circulations were connected. In this coupled system, the PAs acted as pressure sinks. Additionally, existing rotation and hence vorticity of the inflow through the truncal valve favored the development of coherent, streamwise vortices along the PAs comparable to draining or sink vortices \cite{Morton1969}. The low pressure vortex cores affected the flow field downstream of the valve and generated a heterogeneous downstream pressure field. Wider opening of the leaflets aligned with the PAs was facilitated during systole. Furthermore, lower downstream pressures throughout the cycle led to a later pressurization during early diastole and hence later leaflet closure. Eventually, this resulted in asymmetric closure and potentially inadequate leaflet coaptation promoting valve regurgitation. The absence of PAs facilitated a more symmetric closure and adequate leaflet coaptation. Moreover, the remodeling of the vessel geometry towards defined sinuses and a distinct sinotubular junction (STJ) seemed to benefit the wider outward leaflet displacement during systole. 

While TA is classified as total mixing or admixture lesion \cite{Kulik2017}, we identified indicators of favorable streaming of oxygenated blood from the LV toward the aorta and deoxygenated blood from the RV towards the PAs. In the preoperative configuration, a distinct ridge of the FTLE field separated systolic flow from both ventricles. FTLE fields further revealed a clear distinction of blood flow from the RV entering the PAs and blood flow from the LV entering the aorta, resembling a passive separation of pulmonary and systemic circulation. Lagrangian Particle Tracing of `oxygenated' particles seeded in the LV and `deoxygenated' particles in the RV confirmed negligible mixing of both streams until downstream of the valve and mostly deoxygenated flow in the PAs until systolic deceleration. While our analysis was purely based on numerical flow fields and qualitative in its binary categorization of particles, the patient's clinical reports documented exceptionally high values of systemic oxygen saturation for a TA patient. For the first time, we showed favorable streaming in truncus arteriosus and how it ideally benefits the patient. This analysis may facilitate surgical planning for total mixing lesions, preoperatively predicting the favorability of streaming.

In future studies, we plan to parameterize the PA locations to virtually model different types of TA and further assess the impact of their location as it showed to influence hemodynamics, valve performance and streaming of oxygenated and deoxygenated blood. Due to limited image resolution and information about material properties of the neonatal, thickened leaflet tissue, we applied a tested approach for constructing the quadricuspid valve. Based on morphological parameters extracted from echocardiography, this method provides a realistic estimation of the leaflets supporting diastolic pressures. The time-dependent inflow due to ventricular contraction was simulated through a chamber activation model. Incorporating helical ventricular contraction in a flexible ventricle may introduce more complex inflow patterns. To assess streaming, we seeded oxygenated and deoxygenated particles in LV and RV, respectively. As oxygen saturation in the LV is not exactly 100\% and 0\% in the RV, this method provides a qualitative approach which will be extended by further quantitative concentration assessment, e.g. through advection-diffusion modeling. 

In this first study of the fluid mechanics in truncus arteriosus, we investigated the complex flow patterns, vortex dynamics, and mechanisms of mixing between deoxygenated and oxygenated blood in a TA patient modeling the patient-specific FSI between blood flow, truncal valve, and vessel anatomy. Based on comprehensive analysis of the underlying fluid and structural mechanics and their  interaction, we explained mechanisms behind abnormal hemodynamics resulting in inadequate valve performance and showed that the re-establishment of normal hemodynamics improved valve performance after surgical repair of TA. Moreover, we found evidence for beneficial mixing phenomena facilitated by favorable streaming in a total mixing lesion. The versatility of our image-based computational pipeline with patient-specific boundary condition tuning will enable the simulation of any outlet valve disease in future work. This pioneering work on the fluid mechanics of truncus arteriosus will expand the understanding of pathophysiological phenomena and mixing patterns in complex congenital heart disease and total mixing lesions. 


\subsection*{Sources of funding}

KMB and ADK were supported in part by NIH grant K25HL175208, AHA Career Development Award 24CDA1272816 and Stanford Maternal and Child Health Research Institute. PSC was funded by Stanford Maternal and Child Research Institute as the Stephen Bechtel Endowed Fellow in Pediatric Translational Medicine. HLM was funded by NIH grant R38 HL143615. AKR was funded by NIH grant R38 HL143615. ALM and MM were funded by NIH grant R01HL173845.

\subsection*{Disclosures}

None.

\newpage

\begin{figure}[p!]
    \centering
    \includegraphics[width=\textwidth]{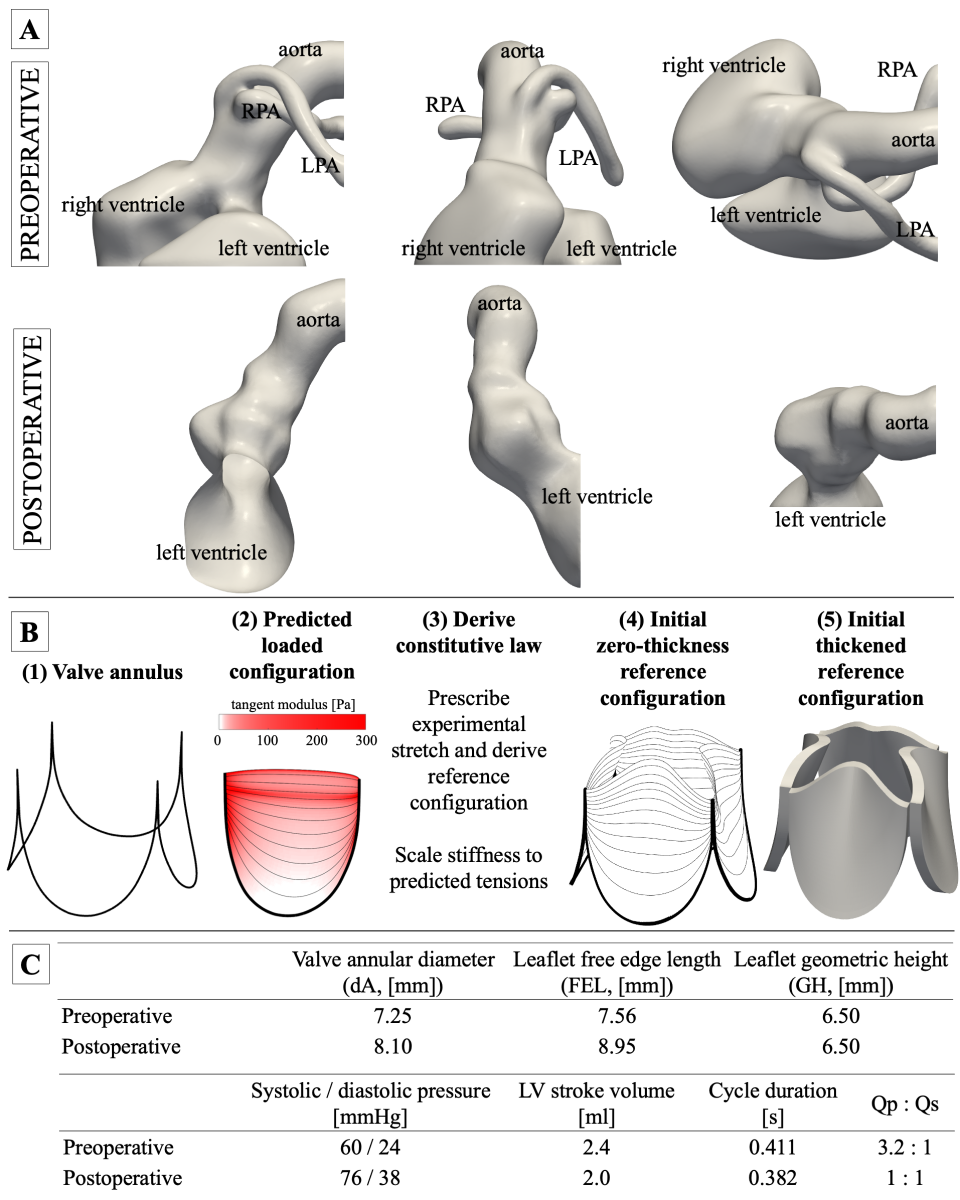}
    \caption{(\textbf{A}) Patient-specific preoperative (top row) and postoperative (bottom row) vessel geometry. Preoperative configuration with left and right ventricle, and left pulmonary artery (LPA) and right pulmonary artery (RPA) branching off the aorta. Postoperative configuration with left ventricle and aorta. Vessel geometries are shown in three views oriented from the same angle each in top and bottom row. (\textbf{B}) Elasticity-based design: (1) Definition of the valve annulus curve. (2) Computation of the predicted loaded configuration. Only one leaflet is shown for visual clarity. (3) Prescription of experimental stretch, scaling of stiffnesses to achieve tension of the predicted loaded configuration, derivation of constitutive law. (4) Definition of the initial zero-thickness configuration under zero pressure. (5) Thickening of the zero-thickness configuration to achieve the valve model used as initial condition in the fluid-structure interaction simulations. (\textbf{C}) Valve morphology parameters and patient-specific hemodynamic parameters.}
    \label{fig:methods}
\end{figure}

\begin{figure}
    \centering
    \includegraphics[width=\textwidth]{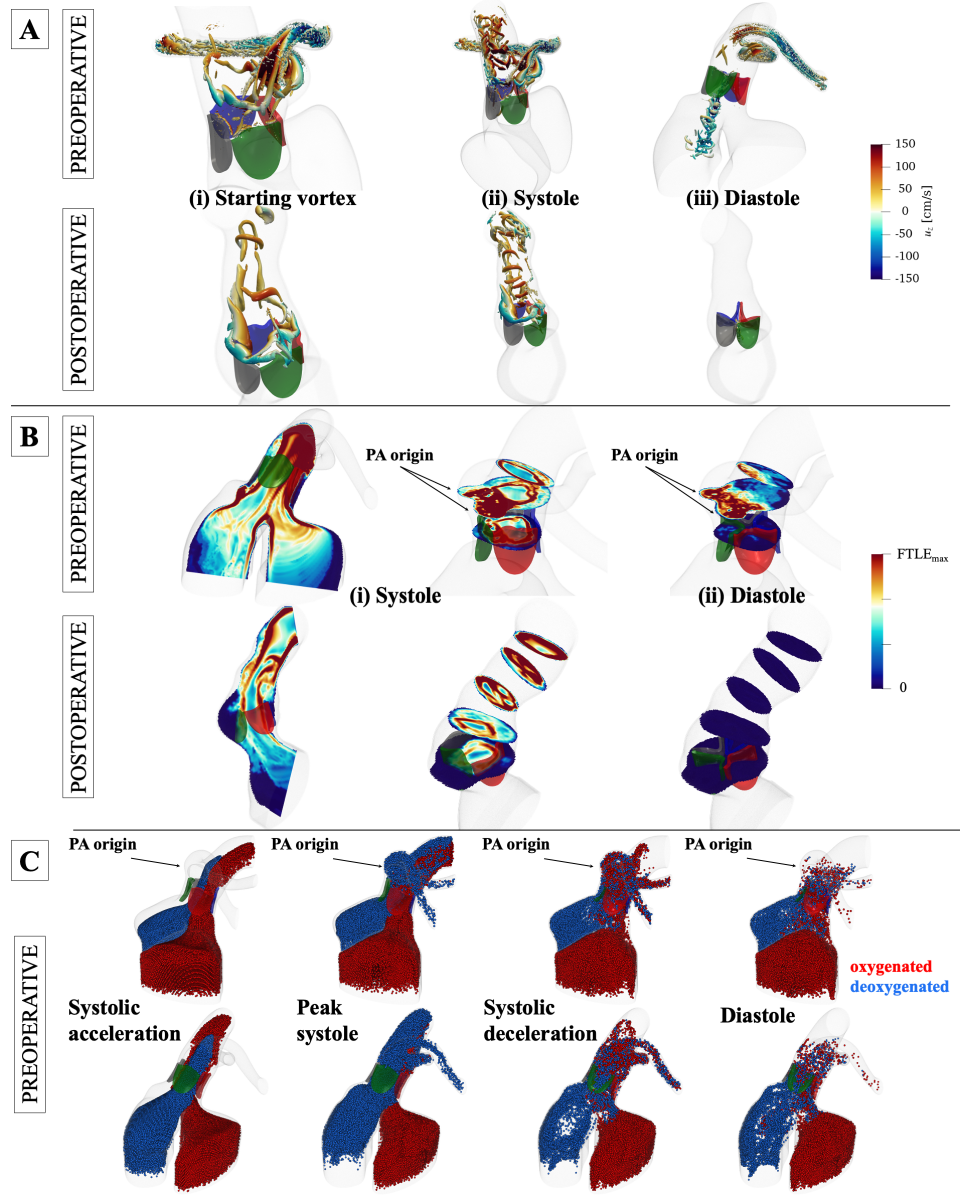}
    \caption{(\textbf{A}) Iso-surfaces of Q-criterion colored by the velocity component normal to valve annular plane for the starting vortex after valve opening, peak systole and mid-diastole in the pre- and postoperative configuration. Quadricuspid valve morphology created a four-lobed starting vortex. Before truncal surgery, streamwise vortices in the PAs influenced the flow field and vortex dynamics in the truncus as they persisted during the entire cardiac cycle. (\textbf{B}) Lagrangian Coherent Structures (LCS): Forward Finite Time Lyapunov Exponent (FTLE) scalar fields for pre- and postoperative configuration. Time instances and scalings are chosen based on relevant flow features. A distinct material surface with negligible mixing was detected along the VSD. LCS revealed upstream effect of the PAs at valve level during systole. During diastole, flow from the outer aortic curvature entered into the PAs. After truncal surgery, systolic hemodynamics were restored with a square-shaped aortic jet due to the quadricuspid valve and negligible diastolic flow during diastole. (\textbf{C}) Lagrangian Particle Tracing: Distribution of `oxygenated' (red) and `deoxygenated' (blue) particles seeded in the left ventricle and right ventricle, respectively, at different view angles and at different time points in the cardiac cycle. Particles are seeded just prior to valve opening and traced over one cardiac cycle. Lagrangian Particle Tracing revealed favorable streaming of oxygenated and deoxygenated flow with streaming from the RV into the PAs and streaming from the LV into the aorta.
    }
    \label{fig:qcrit_FTLE_particle}
\end{figure}

\newpage
  
\bibliographystyle{vancouver}
\bibliography{references}

@article{Griffith2012,
   author = {Boyce E. Griffith},
   doi = {10.1002/cnm.1445},
   issn = {20407947},
   issue = {3},
   journal = {International Journal for Numerical Methods in Biomedical Engineering},
   pages = {317-345},
   pmid = {25830200},
   title = {Immersed boundary model of aortic heart valve dynamics with physiological driving and loading conditions},
   volume = {28},
   year = {2012}
}

@article{Marom2013b,
   author = {Gil Marom and Hee Sun Kim and Moshe Rosenfeld and Ehud Raanani and Rami Haj-Ali},
   doi = {10.1007/s11517-013-1055-4},
   issn = {01400118},
   issue = {8},
   journal = {Medical and Biological Engineering and Computing},
   pages = {839-848},
   pmid = {23475570},
   title = {Fully coupled fluid-structure interaction model of congenital bicuspid aortic valves: Effect of asymmetry on hemodynamics},
   volume = {51},
   year = {2013}
}

@article{Saikrishnan2012,
   author = {Neelakantan Saikrishnan and Choon Hwai Yap and Nicole C. Milligan and Nikolay V. Vasilyev and Ajit P. Yoganathan},
   doi = {10.1007/s10439-012-0527-2},
   isbn = {1043901205272},
   issn = {15739686},
   issue = {8},
   journal = {Annals of Biomedical Engineering},
   pages = {1760-1775},
   pmid = {22318396},
   title = {In vitro characterization of bicuspid aortic valve hemodynamics using particle image velocimetry},
   volume = {40},
   year = {2012}
}

@article{Oechtering2019,
   author = {Thekla H. Oechtering and Malte Sieren and Kathrin Schubert and Tim Schaller and Michael Scharfschwerdt and Apostolos Panagiotopoulos and Buntaro Fujita and Christian Auer and Jörg Barkhausen and Stephan Ensminger and Hans Hinrich Sievers and Alex Frydrychowicz},
   doi = {10.1111/jocs.14253},
   issn = {15408191},
   issue = {12},
   journal = {Journal of Cardiac Surgery},
   pages = {1452-1457},
   pmid = {31638731},
   title = {In vitro 4D Flow MRI evaluation of aortic valve replacements reveals disturbed flow distal to biological but not to mechanical valves},
   volume = {34},
   year = {2019}
}

@article{Gilmanov2018,
   author = {Anvar Gilmanov and Henryk Stolarski and Fotis Sotiropoulos},
   doi = {10.1115/1.4038885},
   issn = {15288951},
   issue = {4},
   journal = {Journal of Biomechanical Engineering},
   pmid = {29305610},
   title = {Flow-Structure Interaction Simulations of the Aortic Heart Valve at Physiologic Conditions: The Role of Tissue Constitutive Model},
   volume = {140},
   year = {2018}
}

@article{Johnson2022,
   author = {Emily L. Johnson and Manoj R. Rajanna and Cheng Hau Yang and Ming Chen Hsu},
   doi = {10.1016/j.finmec.2021.100053},
   issn = {26663597},
   journal = {Forces in Mechanics},
   month = {2},
   pages = {100053},
   publisher = {Elsevier Ltd},
   title = {Effects of membrane and flexural stiffnesses on aortic valve dynamics: Identifying the mechanics of leaflet flutter in thinner biological tissues},
   volume = {6},
   year = {2022}
}

@article{Chen2020,
   author = {Ye Chen and Haoxiang Luo},
   journal = {Journal of Fluid Mechanics},
   pages = {A52},
   publisher = {Cambridge University Press},
   title = {Pressure distribution over the leaflets and effect of bending stiffness on fluid–structure interaction of the aortic valve},
   volume = {883},
   year = {2020}
}

@article{Lee2020,
   author = {Jae H. Lee and Alex D. Rygg and Ebrahim M. Kolahdouz and Simone Rossi and Stephen M. Retta and Nandini Duraiswamy and Lawrence N. Scotten and Brent A. Craven and Boyce E. Griffith},
   doi = {10.1007/s10439-020-02466-4},
   issn = {15739686},
   issue = {5},
   journal = {Annals of Biomedical Engineering},
   pages = {1475-1490},
   pmid = {32034607},
   title = {Fluid–Structure Interaction Models of Bioprosthetic Heart Valve Dynamics in an Experimental Pulse Duplicator},
   volume = {48},
   year = {2020}
}

@article{Bornemann2024_JFM,
   author = {Karoline-Marie Bornemann and Dominik Obrist},
   doi = {10.1017/jfm.2024.309},
   issn = {14697645},
   journal = {Journal of Fluid Mechanics},
   pages = {A41},
   publisher = {Cambridge University Press},
   title = {Instability mechanisms initiating laminar-turbulent transition past bioprosthetic aortic valves},
   volume = {985},
   year = {2024}
}

@article{Fringand2024,
   author = {Tom Fringand and Loic Mace and Isabelle Cheylan and Marien Lenoir and Julien Favier},
   doi = {10.1007/s10439-024-03566-1},
   issn = {0090-6964},
   journal = {Annals of Biomedical Engineering},
   title = {Analysis of Fluid–Structure Interaction Mechanisms for a Native Aortic Valve, Patient-Specific Ozaki Procedure, and a Bioprosthetic Valve},
   year = {2024}
}

@article{Kaiser2019,
   author = {Alexander D. Kaiser and David M. McQueen and Charles S. Peskin},
   doi = {10.1002/cnm.3240},
   issn = {20407947},
   issue = {11},
   journal = {International Journal for Numerical Methods in Biomedical Engineering},
   pmid = {31330567},
   publisher = {Wiley-Blackwell},
   title = {Modeling the mitral valve},
   volume = {35},
   year = {2019}
}

@article{Kaiser2021,
   author = {Alexander D. Kaiser and Rohan Shad and William Hiesinger and Alison L. Marsden},
   doi = {10.1007/s10237-021-01516-7},
   issn = {16177940},
   issue = {6},
   journal = {Biomechanics and Modeling in Mechanobiology},
   pages = {2413-2435},
   pmid = {34549354},
   publisher = {Springer Science and Business Media Deutschland GmbH},
   title = {A design-based model of the aortic valve for fluid-structure interaction},
   volume = {20},
   year = {2021}
}

@article{Naimo2021,
   author = {Phillip S. Naimo and Douglas Bell and Tyson A. Fricke and Yves d'Udekem and Christian P. Brizard and Nelson Alphonso and Igor E. Konstantinov},
   doi = {10.1016/j.jtcvs.2020.04.149},
   issn = {1097685X},
   issue = {1},
   journal = {JTCVS},
   pages = {230-240},
   pmid = {32653289},
   publisher = {Mosby Inc.},
   title = {Truncus arteriosus repair: A 40-year multicenter perspective},
   volume = {161},
   year = {2021}
}

@article{Martinez2019,
   author = {Efrén Martínez-Quintana and Francisco Portela-Torrón},
   doi = {10.21037/TP.2019.02.01},
   issn = {22244344},
   issue = {5},
   journal = {Transl Pediatr},
   pages = {360-362},
   publisher = {AME Publishing Company},
   title = {Truncus arteriosus and truncal valve regurgitation},
   volume = {8},
   year = {2019}
}

@article{Konstantinov2023,
   author = {Igor E. Konstantinov and Christian P. Brizard and Edward Buratto},
   doi = {10.1053/j.pcsu.2022.12.004},
   issn = {18764665},
   journal = {Semin Thorac Cardiovasc Surg Pediatr Card Surg Ann},
   pages = {56-62},
   pmid = {36842799},
   publisher = {W.B. Saunders},
   title = {Congenital Aortic Valve Repair When the Options aren't Good: Truncus Arteriosus and Transposition of the Great Arteries},
   volume = {26},
   year = {2023}
}

@article{Guariento2022,
   author = {Alvise Guariento and Ilias P. Doulamis and Steven J. Staffa and Laura Gellis and Nicholas A. Oh and Takashi Kido and John E. Mayer and Christopher W. Baird and Sitaram M. Emani and David Zurakowski and Pedro J. del Nido and Meena Nathan},
   doi = {10.1016/j.jtcvs.2021.01.136},
   issn = {1097685X},
   issue = {1},
   journal = {JTCVS},
   pages = {224-236},
   pmid = {33726908},
   publisher = {Elsevier Inc.},
   title = {Long-term outcomes of truncus arteriosus repair: A modulated renewal competing risks analysis},
   volume = {163},
   year = {2022}
}

@article{Williams2022,
   author = {Trevor R. Williams and Alana R. Cianciulli and Yan Wang and Andras Lasso and Csaba Pinter and Alison M. Pouch and David M. Biko and Muhammad Nuri and Michael D. Quartermain and Lindsay S. Rogers and Jonathan M. Chen and Matthew A. Jolley},
   doi = {10.1161/CIRCIMAGING.122.014424},
   issn = {19420080},
   issue = {12},
   journal = {Circ Cardiovasc Imaging},
   pages = {E014424},
   pmid = {36093770},
   publisher = {Lippincott Williams and Wilkins},
   title = {Truncal Valve Repair: 3-Dimensional Imaging and Modeling to Enhance Preoperative Surgical Planning},
   volume = {15},
   year = {2022}
}

@article{Bornemann2025_POF,
   author = {Karoline-Marie Bornemann and Dominik Obrist},
   doi = {10.1063/5.0270405},
   issn = {1070-6631},
   issue = {5},
   journal = {Physics of Fluids},
   title = {Leaflet fluttering changes laminar–turbulent transition mechanisms past bioprosthetic aortic valves},
   volume = {37},
   year = {2025}
}

@article{Ferrari2024,
   author = {Lorenzo Ferrari and Dominik Obrist},
   doi = {10.1007/s10439-024-03584-z},
   issn = {15739686},
   journal = {Annals of Biomedical Engineering},
   month = {12},
   publisher = {Springer},
   title = {Comparison of Hemodynamic Performance, Three-Dimensional Flow Fields, and Turbulence Levels for Three Different Heart Valves at Three Different Hemodynamic Conditions},
   year = {2024}
}

@article{Pfaller2021,
   author = {Martin R. Pfaller and Jonathan Pham and Nathan M. Wilson and David W. Parker and Alison L. Marsden},
   doi = {10.1007/s10439-021-02796-x},
   issn = {15739686},
   issue = {12},
   journal = {Annals of Biomedical Engineering},
   month = {12},
   pages = {3574-3592},
   pmid = {34169398},
   publisher = {Springer},
   title = {On the Periodicity of Cardiovascular Fluid Dynamics Simulations},
   volume = {49},
   year = {2021}
}

@article{Kaiser2024,
   author = {Alexander D. Kaiser and Moussa A. Haidar and Perry S. Choi and Amit Sharir and Alison L. Marsden and Michael R. Ma},
   doi = {10.1016/j.jtcvs.2023.12.027},
   issn = {1097685X},
   issue = {3},
   journal = {JTCVS},
   pages = {923-932},
   pmid = {38211896},
   publisher = {Elsevier Inc.},
   title = {Simulation-based design of bicuspidization of the aortic valve},
   volume = {168},
   year = {2024}
}

@article{Simvascular2016,
   author = {Adam Updegrove and Nathan M. Wilson and Jameson Merkow and Hongzhi Lan and Alison L. Marsden and Shawn C. Shadden},
   doi = {10.1007/s10439-016-1762-8},
   issn = {15739686},
   issue = {3},
   journal = {ABME},
   pages = {525-541},
   pmid = {27933407},
   publisher = {Springer New York LLC},
   title = {SimVascular: An Open Source Pipeline for Cardiovascular Simulation},
   volume = {45},
   year = {2017}
}

@article{Naimo2021b,
   author = {Phillip S. Naimo and Tyson A. Fricke and Melissa G.Y. Lee and Yves d'Udekem and Johann Brink and Christian P. Brizard and Igor E. Konstantinov},
   doi = {10.1016/j.jtcvs.2020.01.039},
   issn = {1097685X},
   issue = {2},
   journal = {JTCVS},
   pages = {368-375},
   pmid = {32122575},
   publisher = {Mosby Inc.},
   title = {The quadricuspid truncal valve: Surgical management and outcomes},
   volume = {161},
   year = {2021}
}

@article{Kaiser2025,
   author = {Alexander D. Kaiser and Jing Wang and Aaron L. Brown and Enbo Zhu and Tzung Hsiai and Alison L. Marsden},
   doi = {10.1016/j.jbiomech.2025.112794},
   issn = {00219290},
   journal = {Journal of Biomechanics},
   pages = {112794},
   publisher = {Elsevier BV},
   title = {A fluid-structure interaction model of the zebrafish aortic valve},
   volume = {190},
   year = {2025}
}

@article{Kaiser2023,
   author = {Alexander D. Kaiser and Nicole K. Schiavone and Christopher J. Elkins and Doff B. McElhinney and John K. Eaton and Alison L. Marsden},
   doi = {10.1007/s10439-023-03266-2},
   issn = {15739686},
   issue = {10},
   journal = {Annals of Biomedical Engineering},
   pages = {2267-2288},
   pmid = {37378877},
   publisher = {Springer},
   title = {Comparison of Immersed Boundary Simulations of Heart Valve Hemodynamics Against In Vitro 4D Flow MRI Data},
   volume = {51},
   year = {2023}
}

@article{Choi2024,
   author = {Perry S. Choi and Amit Sharir and Yoshikazu Ono and Masafumi Shibata and Alexander D. Kaiser and Yellappa Palagani and Alison L. Marsden and Michael R. Ma},
   doi = {10.1016/j.xjon.2024.09.008},
   issn = {26662736},
   journal = {JTCVS Open},
   pages = {395-404},
   publisher = {Elsevier B.V.},
   title = {Combined simulation and ex vivo assessment of free-edge length in bicuspidization repair for congenital aortic valve disease},
   volume = {22},
   year = {2024}
}

@article{Griffith2007,
   author = {Boyce E. Griffith and Richard D. Hornung and David M. McQueen and Charles S. Peskin},
   doi = {10.1016/j.jcp.2006.08.019},
   issn = {10902716},
   issue = {1},
   journal = {J. Comput. Phys.},
   month = {4},
   pages = {10-49},
   publisher = {Academic Press Inc.},
   title = {An adaptive, formally second order accurate version of the immersed boundary method},
   volume = {223},
   year = {2007}
}

@article{Choi2025,
   author = {Perry Choi and Amit Sharir and Riya Nilkant and Alexander D Kaiser and Michael Ma},
   city = {Seattle, WA, USA},
   journal = {AATS 105th Annual Meeting},
   title = {Biomechanical Assessment of No-cut Tricuspidization vs Bicuspidization Repairs in Quadricuspid Truncal Valve Model},
   year = {2025}
}

@article{Collett1949,
   author = {Robert W Collett and Jesse E Edwards},
   issue = {4},
   journal = {Surgical Clinics of North America},
   pages = {1245-1270},
   title = {Persistent truncus arteriosus: A classification according to anatomic types},
   volume = {29},
   year = {1949}
}

@article{Bornemann2026_truncal,
   author = {Karoline-Marie Bornemann and Perry S. Choi and Jay Huber and Alexander K. Reed and Amit Sharir and Shiraz A. Maskatia and Alison L. Marsden and Michael R. Ma and Alexander D. Kaiser},
   doi = {10.1016/j.jtcvs.2026.04.039},
   journal = {The Journal of Thoracic and Cardiovascular Surgery},
   title = {Simulations Predict Improved Valve Performance Without Direct Leaflet Intervention After Neonatal Truncus Arteriosus Repair},
   year = {2026}
}

@article{Brown2023,
   author = {Jordan A. Brown and Jae H. Lee and Margaret Anne Smith and David R. Wells and Aaron Barrett and Charles Puelz and John P. Vavalle and Boyce E. Griffith},
   doi = {10.1007/s10439-022-03047-3},
   issn = {15739686},
   issue = {1},
   journal = {Annals of Biomedical Engineering},
   month = {1},
   pages = {103-116},
   pmid = {36264408},
   publisher = {Springer},
   title = {Patient–Specific Immersed Finite Element–Difference Model of Transcatheter Aortic Valve Replacement},
   volume = {51},
   year = {2023}
}

@article{Hurwitz1973,
   author = {Larry E Hurwitz and William C Roberts},
   issue = {5},
   journal = {The American Journal of Cardiology},
   pages = {623-626},
   title = {Quadricuspid Semilunar Valve},
   volume = {31},
   year = {1973}
}

@article{Kaiser2022,
   author = {Alexander D. Kaiser and Rohan Shad and Nicole Schiavone and William Hiesinger and Alison L. Marsden},
   doi = {10.1007/s10439-022-02983-4},
   issn = {15739686},
   issue = {9},
   journal = {Annals of Biomedical Engineering},
   month = {9},
   pages = {1053-1072},
   pmid = {35748961},
   publisher = {Springer},
   title = {Controlled Comparison of Simulated Hemodynamics Across Tricuspid and Bicuspid Aortic Valves},
   volume = {50},
   year = {2022}
}

@article{Yap2010,
   author = {Choon Hwai Yap and Hee-Sun Kim and Kartik Balachandran and Michael Weiler and Rami Haj-Ali and Ajit P Yoganathan},
   doi = {10.1152/ajpheart.00040.2009.-Calcific},
   journal = {Am J Physiol Heart Circ Physiol},
   pages = {395-405},
   title = {Dynamic deformation characteristics of porcine aortic valve leaflet under normal and hypertensive conditions},
   volume = {298},
   year = {2010}
}

@article{May2009,
   author = {Karen May-Newman and Charles Lam and Frank C.P. Yin},
   doi = {10.1115/1.3127261},
   issn = {01480731},
   issue = {8},
   journal = {Journal of Biomechanical Engineering},
   month = {8},
   pmid = {19604021},
   title = {A hyperelastic constitutive law for aortic valve tissue},
   volume = {131},
   year = {2009}
}

@article{Bornemann2026_JFM,
   author = {Karoline Marie Bornemann and Mohammad Moniripiri and Dan S. Henningson and Dominik Obrist and Peter J. Schmid and Ardeshir Hanifi},
   doi = {10.1017/jfm.2026.11350},
   issn = {14697645},
   journal = {Journal of Fluid Mechanics},
   month = {3},
   publisher = {Cambridge University Press},
   title = {Optimal three-dimensional perturbations in fluttering and non-fluttering bioprosthetic aortic valves},
   volume = {1031},
   year = {2026}
}

@article{Morton1969,
   author = {B. R. Morton},
   doi = {10.1017/S002211206900019X},
   issn = {14697645},
   issue = {2},
   journal = {Journal of Fluid Mechanics},
   month = {9},
   pages = {315-333},
   title = {The strength of vortex and swirling core flows},
   volume = {38},
   year = {1969}
}

@article{Vismara2014,
   author = {Riccardo Vismara and Andrea Mangini and Claudia Romagnoni and Monica Contino and Alberto Redaelli and Gianfranco B. Fiore and Carlo Antona},
   journal = {The Journal of Heart Valve Disease},
   pages = {122-126},
   title = {In-Vitro Study of a Porcine Quadricuspid Aortic Valve},
   volume = {23},
   year = {2014}
}

@article{Lasso2022,
   author = {Andras Lasso and Christian Herz and Hannah Nam and Alana R. Cianciulli and Steve Pieper and Simon Drouin and Csaba Pinter and Samuelle St-Onge and Chad Vigil and Stephen Ching and Kyle Sunderland and Gabor Fichtinger and Ron Kikinis and Matthew A. Jolley},
   journal = {Frontiers in Cardiovascular Medicine},
   pages = {886549},
   title = {SlicerHeart: An open-source computing platform for cardiac image analysis and modeling},
   volume = {9},
   url = {https://github.com/},
   year = {2022}
}

@article{Lavon2018,
   author = {Karin Lavon and Rotem Halevi and Gil Marom and Sagit Ben Zekry and Ashraf Hamdan and Hans Joachim Schäfers and Ehud Raanani and Rami Haj-Ali},
   doi = {10.1115/1.4038329},
   issn = {15288951},
   issue = {3},
   journal = {Journal of Biomechanical Engineering},
   month = {3},
   pmid = {29098290},
   publisher = {American Society of Mechanical Engineers (ASME)},
   title = {Fluid-Structure Interaction Models of Bicuspid Aortic Valves: The Effects of Nonfused Cusp Angles},
   volume = {140},
   year = {2018}
}

@inbook{Kulik2017,
   author = {Thomas J Kulik},
   edition = {5},
   editor = {Richard A. Polin and Steven H. Abman and David H. Rowitch and William E. Benitz and William W. Fox},
   booktitle = {Fetal and neonatal physiology},
   publisher = {Elsevier},
   title = {Physiology of Congenital Heart Disease in the Neonate},
   year = {2017}
}

@article{Schrauben2019,
   author = {Eric M. Schrauben and Brahmdeep Singh Saini and Jack R.T. Darby and Jia Yin Soo and Mitchell C. Lock and Elaine Stirrat and Greg Stortz and John G. Sled and Janna L. Morrison and Mike Seed and Christopher K. MacGowan},
   doi = {10.1186/s12968-018-0512-5},
   issn = {1532429X},
   issue = {1},
   journal = {Journal of Cardiovascular Magnetic Resonance},
   month = {1},
   pmid = {30661506},
   publisher = {BioMed Central Ltd},
   title = {Fetal hemodynamics and cardiac streaming assessed by 4D flow cardiovascular magnetic resonance in fetal sheep},
   volume = {21},
   year = {2019}
}

@article{Shadden2005,
   author = {Shawn C. Shadden and Francois Lekien and Jerrold E. Marsden},
   doi = {10.1016/j.physd.2005.10.007},
   issn = {01672789},
   issue = {3-4},
   journal = {Physica D: Nonlinear Phenomena},
   month = {12},
   pages = {271-304},
   publisher = {Elsevier},
   title = {Definition and properties of Lagrangian coherent structures from finite-time Lyapunov exponents in two-dimensional aperiodic flows},
   volume = {212},
   year = {2005}
}

@article{Chen2016,
    title = {{Outcomes of surgical repair for persistent truncus arteriosus from neonates to adults: A single center's experience}},
    year = {2016},
    journal = {PLoS ONE},
    author = {Chen, Qiuming and Gao, Huawei and Hua, Zhongdong and Yang, Keming and Yan, Jun and Zhang, Hao and Ma, Kai and Zhang, Sen and Qi, Lei and Li, Shoujun},
    number = {1},
    month = {1},
    volume = {11},
    publisher = {Public Library of Science},
    doi = {10.1371/journal.pone.0146800},
    issn = {19326203},
    pmid = {26752522}
}

@article{Naimo2018,
    title = {{Impact of truncal valve surgery on the outcomes of the truncus arteriosus repair}},
    year = {2018},
    journal = {European Journal of Cardio-thoracic Surgery},
    author = {Naimo, Phillip S. and Fricke, Tyson A. and D'Udekem, Yves and Brink, Johann and Weintraub, Robert G. and Brizard, Christian P. and Konstantinov, Igor E.},
    number = {3},
    month = {9},
    pages = {524--531},
    volume = {54},
    publisher = {European Association for Cardio-Thoracic Surgery},
    doi = {10.1093/ejcts/ezy080},
    issn = {1873734X},
    pmid = {29528381}
}

@misc{Wong2014,
    title = {{Streaming in transposition of the great arteries by using cardiac magnetic resonance imaging}},
    year = {2014},
    booktitle = {Circulation},
    author = {Wong, James and Pushparajah, Kuberan and Hussain, Tarique and Giese, Daniel and Dedieu, Nathalie and Mathur, Sujeev and Greil, Gerald F. and Razavi, Reza and Bell, Aaron},
    number = {10},
    month = {3},
    pages = {1169--1170},
    volume = {129},
    doi = {10.1161/CIRCULATIONAHA.113.002852},
    issn = {00097322},
    pmid = {24615966}
}

\newpage

\section*{Supplementary Material}
\label{sec:supp}

\section*{Detailed Methods}
Additional information on the computational methods is provided below.

\subsection*{Governing equations}
\label{subsec:governing_equations}

The interaction of the valve leaflets and vessel wall with the surrounding blood flow was modeled via the Immersed Boundary Method implemented in the software library IBAMR (Immersed Boundary Adaptive Mesh Refinement \cite{Griffith2007}). Coupling between the Eulerian frame with a staggered Cartesian grid (fluid domain) and the Lagrangian frame with an unstructured mesh (structural domain: vessel wall and valve) was achieved via convolutions with a Dirac $\delta$ function. Structural inlets and outlets were aligned with the fluid grid boundaries and a fixed position of the structural domain was enforced by target points via a penalty method 
\begin{equation}
    \mathbf{F} = -k (\mathbf{X} - \mathbf{X}_{target})
\end{equation}
which corresponds to a linear spring with zero rest length between the current and desired position. The fluid was modeled as incompressible with a density of $\rho = 1.0 \, g/cm^3$ and a dynamic viscosity of $\mu = 0.04\, Poise$. The governing equations are defined as
\begin{align}
    \rho\left(\frac{\partial \mathbf{u}\left(\mathbf{x},t\right)}{\partial t} + \mathbf{u}\left(\mathbf{x},t\right) \cdot \nabla \mathbf{u}\left(\mathbf{x},t\right)\right) &= -\nabla p\left(\mathbf{x},t\right) + \mu \Delta \mathbf{u}\left(\mathbf{x},t\right) + \mathbf{f}\left(\mathbf{x},t\right) \label{equ:momentum} \\
    \nabla \cdot \mathbf{u} \left(\mathbf{x},t\right) &= 0 \label{equ:continuum} \\
    \mathbf{F}\left(\cdot, t\right) &= \mathcal{F}\left(\mathbf{X}\left(\cdot,t\right)\right) \label{equ:force_map} \\
    \frac{\partial \mathbf{X}\left(\mathbf{s},t\right)}{\partial t} &= \mathbf{u}\left(\mathbf{X}\left(\mathbf{s},t\right),t\right) \nonumber \\
    &= \int \mathbf{u}\left(\mathbf{x},t\right)\delta \left(\mathbf{x} - \mathbf{X}\left(\mathbf{s},t\right)\right) \text{d}\mathbf{x} \label{equ:interaction}\\
    \mathbf{f}\left(\mathbf{x},t\right) &= \int \mathbf{F}\left(\mathbf{s},t\right) \delta \left(\mathbf{x} - \mathbf{X}\left(\mathbf{s},t\right)\right)\text{d}\mathbf{s} \quad . 
    \label{equ:force_spread}
\end{align}
Equations \ref{equ:momentum} and \ref{equ:continuum} are the incompressible Navier-Stokes equations describing the conservation of mass and momentum with fluid velocity $\mathbf{u}$, pressure $p$, fluid density $\rho$ and fluid dynamic viscosity $\mu$. Equation \ref{equ:momentum} also includes the structure force $\mathbf{f}(\mathbf{x},t)$ acting on the fluid at fixed Eulerian spatial positions $(\mathbf{x},t)$. Equation \ref{equ:force_map} describes the mapping of the entire structural configuration $\mathcal{F}(\mathbf{X}(\cdot,t))$ at time $t$ to the Lagrangian force $\mathbf{F}(\cdot, t)$ and the current structural configuration $\mathbf{X}(\mathbf{s},t)$ where $\mathbf{s}$ labels material points. Equations \ref{equ:interaction} and \ref{equ:force_spread} couple the fluid and structure motion via the Dirac delta function $\delta$. Equation \ref{equ:interaction} states that the structure moves with the fluid velocity while equation \ref{equ:force_spread} converts $\mathbf{F}$ to the structure force acting on the fluid $\mathbf{f}$. 

Both pre- and postoperative fluid domains were discretized by a mesh width of $0.1\,mm$ to achieve adequate resolution within the PAs with a diameter of $2\,mm$. This led to $360 \times 320 \times 248$ points for the preoperative fluid domain and $272 \times 240 \times 416$ points for the postoperative fluid domain. The structural mesh width was targeted to $3/4$ of the fluid mesh width ($\Delta s = 3 \Delta x/4 = 0.075\,mm$). The time step size was $\Delta t = 3.0 \cdot 10^{-6}\, s$.

To ensure adequate resolution of transitional flow structures during systole, we calculated an approximate Kolmogorov scale for pre- and postoperative configuration, which corresponds to $300\mu m$ and $470\mu m$, respectively. This results in a ratio between Kolmogorov scale and fluid mesh width of $3$ and $4.7$, respectively, which confirms that we fully resolve turbulent structures in the flow field.

Three cycles were simulated of which the second cycle was used for analysis. Simulations were performed on Stanford University's Sherlock cluster on $48$ Intel Xeon Gold $5118$ cores across $2$ nodes with $2.30$GHz clock speed. 

\subsection*{Simulation setup}
\label{subsec:simulation_setup}

\subsubsection*{Patient selection}
\label{subsubsec:patient_selection}

The neonatal patient was selected based on the availability of pre- and postoperative imaging and the absence of any surgical modification of the quadricuspid truncal valve. The patient was diagnosed with quadricuspid valve morphology of Hurwitz and Roberts type A with four symmetric leaflets \cite{Hurwitz1973} and truncus arteriosus of Collett and Edwards type II with close proximity of the pulmonary artery origins at a single circumferential location \cite{Collett1949}. Preoperative CT imaging was obtained at 2 days of age while postoperative CT imaging was recorded at 1 month of age. 

\subsubsection*{Image segmentation of the vessel geometry}
\label{subsubsec:image_segmentation}

Segmentation of CT imaging scans was performed in the open-source software SimVascular \citep{Simvascular2016}. After identifying centerlines of each vessel, two-dimensional contours were created to generate a lofted geometry described by a triangular surface mesh. The surface mesh was then thickened by two additional layers to achieve a wall thickness of two times the structural mesh width \cite{Kaiser2019}. An approximately constant spatial position of the vessel was enforced by target points and the layers were held together by linear springs. Pre- and postoperative vessel geometries were shown in Figure \ref{fig:methods}A. The preoperative configuration partially includes LV and RV with a VSD proximal to the quadricuspid truncal valve. Distal to the truncal valve, the left (LPA) and right pulmonary arteries (RPA) branched from the right-sided aorta in very close proximity approximately at the STJ. The postoperative configuration was composed of the partial LV proximal to the valve, now bounded by a surgically closed VSD, and the right-sided aortic arch distal to the valve. Both vessel geometries were truncated at selected locations and flow extenders were placed at each inlet and outlet to align the structural domain with the Cartesian boundaries of the fluid domain. 

\subsubsection*{Quadricuspid valve construction: Elasticity-based design} 
\label{subsubsec:quadricuspid_construction}

The quadricuspid valve was constructed based on an elasticity-based design approach introduced by Kaiser and colleagues for the mitral \cite{Kaiser2019} and aortic valve \cite{Kaiser2021}. This framework is applicable to a wide range of scales \cite{Kaiser2025} and showed excellent in vitro \cite{Kaiser2023}, ex vivo \cite{Choi2024} and in vivo agreement with patient imaging data \cite{Bornemann2026_truncal}. Using this methodology, the heterogeneous leaflet fiber structure and material were defined via tuning parameters in the leaflet equilibrium equations under the condition that the valve supports a given pressure (Figure \ref{fig:methods}B). First, the valve annulus was defined (Figure \ref{fig:methods}B,1). Each leaflet was represented as an unknown parametric surface in $\mathbb{R}^3$ with the spatial location $\mathbf{X}$ defined by coordinates $u$ and $v$ aligned with the circumferential and radial material directions, respectively. The circumferential direction corresponded to fiber direction while the radial direction represented the cross-fiber direction. The mechanical equilibrium of the valve leaflet was given by
\begin{equation}\label{equ:equilibrium}
    0 = p\left(\mathbf{X}_u \times \mathbf{X}_v\right) + \frac{\partial}{\partial u}\left(S \frac{\mathbf{X}_u}{|\mathbf{X}_u|}\right) + \frac{\partial}{\partial v}\left(T \frac{\mathbf{X}_v}{|\mathbf{X}_v|}\right)
\end{equation}
with $p$ as the pressure supported by the leaflet and $\mathbf{X}_u$ and $\mathbf{X}_v$ as partial derivatives in circumferential and radial directions, respectively. Unit tangents are described as $\mathbf{X}_u/|\mathbf{X}_u|$ in circumferential direction and $\mathbf{X}_v/|\mathbf{X}_v|$ in radial direction. Variables $S$ and $T$ state the tension exerted by the leaflet in circumferential and radial direction, respectively. To close this system of equations, the tensions exerted by the leaflet in circumferential direction ($S$) and radial direction ($T$) are temporarily defined as 
\begin{align}\label{equ:tensions}
\begin{split}
    S\left(u,v\right) &= \alpha \left(1-\frac{1}{1+|\mathbf{X}_u|^2/a^2}\right) \\
    T\left(u,v\right) &= \beta \left(1-\frac{1}{1+|\mathbf{X}_v|^2/b^2}\right)
\end{split}
\end{align}
with tunable and not necessarily constant parameters $\alpha,\beta,a,b$. Parameters $\alpha$ and $\beta$ state the maximum tension reachable by each fiber and parameters $a$ and $b$ allow control of fiber spacing to achieve the desired valve morphology during closure. 

Equations \ref{equ:tensions} were incorporated into equation \ref{equ:equilibrium} and the resulting system was discretized via centered finite differences. The solution of this nonlinear system predicted a closed, loaded leaflet configuration including leaflet geometry and fiber orientations (Figure \ref{fig:methods}B,2). 

From this configuration, the constitutive law and reference configuration of the valve were derived via membrane stiffnesses to achieve the predicted tensions at a known experimentally derived stretch (Figure \ref{fig:methods}B,3). First, we imposed an in vitro measured uniform circumferential stretch ratio ($\lambda_c = 1.15$) and radial stretch ratio ($\lambda_r=1.54$) \cite{Yap2010}. Next, we solved for the reference length $R$ in $\lambda = L/R$ for the respective fiber link directions. The tension/stretch relation is exponential with rates of $\eta_c = 57.46$ circumferentially and $\eta_r = 22.40$ radially \cite{May2009}. We then solved the following relation for $\kappa$ as local stiffness coefficient to scale the stiffness of each link to match the predicted tension $\tau$ 
\begin{equation}
    \tau = \kappa (e^{\eta (\lambda-1)} -1) \quad .
\end{equation}
As stiffness coefficients are uniquely assigned to each link, we achieved a heterogeneous distribution of leaflet stiffness in the reference configuration. 

Next, we used the obtained reference configuration to generate an open, relaxed leaflet configuration serving as initial condition for the following fluid-structure interaction simulations (Figure \ref{fig:methods}B,4). Applying the derived constitutive law for the tension distribution, the leaflet mechanical equilibrium (equation \ref{equ:equilibrium}) was solved again, but now prescribing $p=0$. This way, we obtained an open configuration, which maintains reduced but non-zero pre-strain and residual tension, as compared to the predicted loaded configuration. 

As a last step, the derived zero-thickness leaflet configuration was thickened by four additional layers in its normal direction (Figure \ref{fig:methods}B,5). The resulting layers were connected by linear springs and each layer was assigned an individual membrane stiffness with the total stiffness divided by 5. In both pre- and postoperative configuration, we assigned a nominal uniform thickness of $0.5\,mm$ as leaflet thickening was observed in echocardiography imaging. 

The selected patient showed four nearly symmetric leaflets, corresponding to a quadricuspid valve type A \cite{Hurwitz1973}. Patient-specific free edge length (FEL) and geometric height (GH) were measured from echocardiography imaging before and after TA repair (Figure \ref{fig:methods}C) and the gross valve morphology was matched via elasticity-based design \cite{Kaiser2021}. In both configurations, the four leaflets were attached to a rigid scaffold which was aligned with the commissures based on CT imaging. Location and orientation of the valve were confirmed and validated by an expert imaging cardiologist (SAM). 

\subsubsection*{Derivation of patient-specific boundary conditions}
\label{subsubsec:boundary_conditions}

Patient-specific boundary conditions were derived from clinically measured patient vital signs, CT and echocardiography imaging. Due to the sparse nature of hemodynamic measurements in neonates, we additionally derived parameters based on physiological concepts and clinical literature. To match patient-specific boundary conditions, we performed multiple iterations between high-fidelity, three-dimensional FSI simulations and the equivalent zero-dimensional configuration using svZeroDSolver implemented in SimVascular \cite{Pfaller2021}. 

Before TA repair, the computational domain consisted of LV and RV as inlets and aorta, LPA, and RPA as outlets. The patient showed a heart rate of $146\, bpm$ corresponding to a cardiac cycle duration of $0.411\,s$. For inflow modeling, we used a time-varying elastance chamber model \cite{Brown2023} targeting a patient-derived LV stroke volume of $2.4\,ml$. Based on the assumption that the pressure was identical in LV and RV due to an unrestricted VSD, we impose the LV pressure at the RV inlet as a pressure boundary condition. At the outlets, we aimed to match a pulmonary (Qp) to systemic (Qs) flow ratio between $3$ to $1$ and $4$ to $1$, as calculated based on the patient's systemic oxygen saturation. In our simulation, we achieved a $3.2$ to $1$ Qp:Qs ratio by tuning the resistance ratio imposed via Resistance-Capacitance-Resistance (RCR) boundary conditions at the aortic and pulmonary outlets. Simultaneously, we targeted patient-specific systolic to diastolic pressures of $60/24\,mmHg$ at the aortic outlet. For the postoperative configuration, the patient's heart rate was $157\,bpm$ resulting in a cardiac cycle of $0.382\,s$. We tuned the time-varying elastance chamber model for the LV and the outlet RCR boundary condition to achieve a stroke volume of $2.0\,ml$ and systolic/diastolic pressures of $76/38\,mmHg$ at the aortic outlet (Figure \ref{fig:methods}C). 

\subsection*{Hemodynamics, waveforms and leaflet kinematics}
\label{subsec:method_hemo_kinematics}
To assess the flow field in pre- and postoperative configuration, vessel centerlines were extracted using 3DSlicer \cite{Lasso2022}. Perpendicular slices were created at selected locations excluding the junctions and the streamwise velocity component $\mathbf{u}_s$ was computed normal to each slice. Flow and pressure waveforms were extracted at the inlets (LV, RV) and outlets (aorta, LPA, RPA) as well as at a slice located $0.12$ cm above the valve commissure level which corresponds to the region upstream of the PAs in the preoperative configuration.

To evaluate leaflet kinematics, the location of the central leaflet tip was tracked over time. The time-dependent displacement was calculated based on the Euclidean distance between the current leaflet tip position and the central coaptation point during diastole. 

\subsection*{Vortex development and breakdown}
\label{subsec:method_LCS}

Vortex development and breakdown were assessed by iso-surfaces of the Q-criterion colored by the velocity component normal to the valve annular plane. 

\subsection*{Lagrangian Coherent Structures}
\label{subsec:method_LCS}

Lagrangian Coherent Structures (LCS) are second-derivative ridges or valleys of the scalar Finite Time Lyapunov Exponent (FTLE) field \cite{Shadden2005}. The FTLE field $\sigma_t^T(\mathbf{x})$ defines a finite time average of the maximum separation of a particle pair advected with the flow. Based on the maximum eigenvalue of the right Cauchy-Green deformation tensor $\lambda_{\text{max}}$ of the Lagrangian flow map differentiated with respect to material point labels associated with the point $\mathbf{x}$, the maximum FTLE is defined as 
\begin{equation}
    FTLE = \sigma_{t_0}^T = \frac{1}{|T|}\text{ln}\sqrt{\lambda_{\text{max}}(\mathbf{x})} \quad .
\end{equation}
The integration time was taken to be positive. Forward-time integration results in repelling LCS (stable manifolds of a time-dependent vector field). Flux across these invariant manifolds is negligible, meaning that they can be interpreted as transport barriers. Since these barriers are advected with the flow, we describe them henceforth as material surfaces \cite{Shadden2005}. 

\subsection*{Lagrangian Particle Tracing}
\label{subsec:method_particle}

Following Lagrangian Coherent Structures, we qualitatively investigated the mixing of oxygenated and deoxygenated blood via mass-less Lagrangian particles, which were seeded within the ventricles sufficiently proximal to the valve. Particles within the LV were assumed as `oxygenated', hence colored in red. Particles within the RV were assumed as `deoxygenated', hence colored in blue. Figure \ref{fig:qcrit_FTLE_particle} shows the Lagrangian particle distribution at different views and at different instances during the cardiac cycle. Particles were released just prior to the valve opening and traced over the second cardiac cycle.

\end{document}